\documentclass{amsart}
\usepackage{graphicx}
\usepackage{amscd}
\usepackage{thmdefs}

\input{tcilatex}
\theoremstyle{definition}
\theoremstyle{remark}
\numberwithin{equation}{section}

\input tcilatex

\begin{document}
\author{Miguel Tierz}
\title{Quantum group symmetry and discrete scale invariance: Spectral aspects}
\address{Institut d'Estudis Espacials de Catalunya (IEEC), Campus UAB, Fac. Ciencies,
Torre C5-Par-2a pl, E-08193 Bellaterra (Barcelona), Spain\\
tierz@ieec.fcr.es}
\address{The Open University, Applied Mathematics Department., Milton Keynes, MK7
6AA, UK\\
M.Tierz@open.ac.uk}
\date{}
\maketitle

\begin{abstract}
We study analytical aspects of a generic $q$-deformation with $q$ real, by
relating it with discrete scale invariance. We show how models of conformal
quantum mechanics, in the strong coupling regime and after regularization,
are also discrete scale invariant. We discuss the consequences of their
distinctive spectra, characterized by functional behavior. The role of
log-periodic behavior and $q$-periodic functions is examined, and we show
how $q$-deformed zeta functions, characterized by complex poles, appear. As 
an application, we discuss one-loop effects in discretely self-similar
space-times.
\end{abstract}

\section{Introduction}

The number of physical models that lead to a quantum mechanical spectrum
with functional behavior is remarkable \cite{Camblong:2000ec}--\cite
{LeClair:2002ux}. However, this is not a very well-known fact and indeed,
most of the works that posses at least this common feature appear rather
scattered in the literature. By functional behavior we mainly refer here to
spectra with exponential (either exactly or exponential in the semiclassical
region) growth. The usual models of ordinary quantum mechanics certainly
seem to imply that only polynomial or algebraic behavior in the quantum
numbers can be obtained.

This type of spectral behavior goes hand in hand with the presence of a
quantum group symmetry. Indeed, in ordinary quantum mechanics, when $q$%
-deformed, the usual polynomial behavior of the spectrum jumps into an
exponential-like behavior for the eigenvalues \cite{Macfarlane:dt,Majid:kd}.
Is possible to obtain this type of spectra within ordinary quantum mechanics
? The answer turns out to be positive, but for a rather particular type of
models. More precisely, models of conformal quantum mechanics \cite
{deAlfaro:1976je}. These are models that are characterized by a singular
potential \cite{Case}, like $V\left( r\right) =-\frac{\lambda }{r^{2}},$ for
example. These models have been notoriously revisited in recent years \cite
{Camblong:2000ec,Camblong:2003mz}. Their interest partially relies on the
fact that they are a good laboratory to test many quantum field theory
features (like regularization and renormalization), but also due to its
physical relevance in many different areas, as we shall see. Qualitatively,
one may argue that the dimensionless \footnote{%
Due to the concomitant symmetry of the potential, of degree $-2$, with the
kinetic energy term.} parameter $\lambda $, is willing to play the role of
the dimensionless $q$ (rather, $\log q)$ parameter$.$ Nevertheless, this is
not possible since we do not have an energy scale, given by a more
conventional dimensionful parameter. This parameter is given by the usual
and necessary cutoff employed in the regularization (or appearing by
dimensional transmutation \cite{Camblong:2003mz} for example). Then, this
new parameter behaves like a usual parameter in ordinary quantum mechanics
and then $\lambda $ is legitimated to act as an \textit{honest} $q$
parameter. At any rate, the resulting spectrum: 
\begin{equation}
E_{n}=E_{0}\mathrm{e}^{-n\mu }, \label{exp}
\end{equation}
with $E_{0}$ depending on the cutoff and $\mu $ depending on $\lambda $ is
rather conclusive and certainly close to some typical $q$-deformed spectra 
\cite{Skorik:1993ce}.

A feature of $q$ deformed models that we want to stress is their
relationship with the concept of discrete-scale invariance \cite{Sor}.
Discrete scale invariance (DSI) is a symmetry that is weaker than the
well-known continuous scale invariance. In discrete scale invariance, we
have scale invariance under a generic transformation $x\rightarrow \lambda x$
only for specific values of the parameter $\lambda $. In general, these
values form an infinite but countable set that can be expressed as $\lambda
_{n}=\lambda ^{n}$, with $\lambda $ playing the role of a fundamental
scaling ratio. Regarding $q$-deformed models, is not difficult to realize
that they are DSI invariant with the parameter $q$ playing the role of $%
\lambda .$ In addition, since continuous scale invariance is equivalent to
continuous translational invariance expressed on the logarithms of the
variables \cite{Sor}, then DSI can also be considered as translational
invariance, but restricted to a discrete set. That is, only valid for
translations of the fundamental unit $\log \lambda .$ This is essentially
the well-known \textit{lattice} structure associated to $q$-deformations
(see \cite{Fichtmuller:1995dt} for example).

Already in the seventies, in nuclear physics work -more precisely, the
Efimov effect \cite{Efimov}-, we find evidence of the physical relevance of
a quantum mechanical spectrum with an exponential behavior. Also in recent
work \cite{Glazek:1993rc}, there is emphasis in a spectrum of the type $%
\left( \ref{exp}\right) ,$ and also its relationship with limit cycles in
renormalization group flows \cite{Wilson:1970ag}-\cite{Morozov:2003ik} and
under which conditions the spectrum is of pure geometric growth. Needless to
say, discrete scale invariance might lead to any kind of functional behavior
for the spectrum and another, closely related, possibility is:
\begin{equation}
E_{n}=\frac{E_{0}}{\sinh \left( n\alpha \right) }, \label{sinh}
\end{equation}
that also appears, for example, in \cite{LeClair:2002ux}, and is probably
somewhat closer to $q$-deformed models than $\left( \ref{exp}\right) $. As
we shall see, this spectrum, with linear growth near the ground-state and
exponential one for high-lying eigenvalues, presents some interesting
features when compared to $\left( \ref{exp}\right) .$

The paper is organized as follows. In the next Section we discuss how
quantum mechanical models with conformal symmetry posses, after
regularization, common properties with $q$-deformed models. Then, in Section
3, we study some spectral functions -like the partition function or the
density of states- associated with a purely exponential spectrum. We discuss
the role of the complex poles and when these poles induce fractal behavior.
Then, we proceed identically, but for a typical $q$-deformed spectrum. For
this, we employ some mathematical works on $q$-deformed zeta functions, and
we see that there is a considerably richer oscillatory behavior, due to an
interesting meromorphic structure of the Mellin transforms. In the last
Section, we study one-loop effects in discretely self-similar space-times,
that both shows a physical application of $q$-deformed zeta functions and
also exploits the connections between quantum group symmetries and discrete
scale invariance. In the Conclusions, we present a brief summary and some
avenues for future research are suggested.

\section{Regularization in conformal quantum mechanics and quantum group
symmetries}

As we have mentioned in the Introduction, quantum mechanical models with
conformal symmetry, characterized by singular potentials, posses several
rather special features. We shall mainly exemplify our discussion with the
potential $V\left( r\right) =-\frac{\lambda }{r^{2}}.$ As already mentioned,
the parameter $\lambda $ is, in contrast to ordinary quantum mechanics,
adimensional, due to the concomitant symmetry with the kinetic term. Note
that this leads to conformal symmetry. Actually, it is imposed by it.

In this model, big values of the parameter $\lambda $ are known to lead to
the so-called strong coupling regime where regularization and
renormalization are mandatory. This may be done in several ways, as has been
discussed in detail \cite{Camblong:2000ec,Camblong:2003mz}. In any case,
either by the presence of a dimensional cutoff or by dimensional
transmutation for example, one finally ends up with an additional
dimensional parameter. This parameter gives an energy scale and the result
is that it allows. There are many features of the model that puts into
evidence the discrete scale invariance. Note that in $q$-deformations one
typically begins with ordinary quantum mechanics, with a dimensional
parameter (like $\omega $ in a harmonic oscillator), and then the
generalization of the algebra is achieved with the introduction of an
adimensional parameter $q.$ Roughly speaking, the final situation is the
same but the steps are done in opposite order.

However, it must be stressed that while the introduction of a regularization
parameter leads to an exponential spectrum $\left( \ref{exp}\right) ,$ a
posterior renormalization \cite{CoonHol} eliminates the excited states and
leaves only the ground state.$.$ However, since regularization of the
singular and, eventually, non-physical part of the potential is a meaningful
procedure in many physical applications and in any case, this type of
spectra explicitly appears in physical problems, we shall be studying some
of its consequences and how they lead to log-periodic behavior, the
signature of discrete scale invariance.

Note also that the \textit{intriguing} \cite{Camblong:2003mz} oscillatory
behavior associated to this model in the strong-coupling regime, as
exemplified through its wavefunction is again easily understood and even
expected in terms of discrete scale invariance. For $r>a$ and energy $E<0$ 
\cite{Camblong:2003mz,Case}, 
\begin{eqnarray}
v^{(>)}(r)\!\! &=&A_{l,\nu }\,K_{i\Theta }(\kappa r)  \label{eq:ISP_BS_wf_>}
\\
\!\! &\overset{(a<r\rightarrow 0)}{=}&\!\!-A_{l,\nu }\;\sqrt{\frac{\pi }{%
\Theta \sinh \left( \pi \Theta \right) }}\,  \notag \\
&\times &\!\!\!\!\!\!\!\!\!\!\!\!\!\!\!\!\sin \left\{ \Theta \left[ \ln
\left( \frac{\kappa r}{2}\right) +\gamma \right] \right\} \!\left[ 1+O\left(
[\kappa r]^{2}\right) \right] \,,  \label{eq:MacDonald_asymptotic}
\end{eqnarray}
where $K_{i\Theta }(z)$ is the Macdonald function of imaginary order $%
i\Theta $, $\gamma $ is the Euler-Mascheroni constant, and 
\begin{equation}
\Theta =\sqrt{\lambda -(l+\nu )^{2}}\;,  \label{eq:Theta_coupling}
\end{equation}
while $\nu =d/2-1$, $\lambda =2mg/\hbar ^{2}$, and $\kappa ^{2}=-2mE/\hbar
^{2}$. In ~(\ref{eq:ISP_BS_wf_>})-(\ref{eq:MacDonald_asymptotic}) and
thereafter, the reduced function $v(r)$ in $d$ dimensions is defined in
terms of the separable solution $\Psi (\mathbf{r})\propto Y_{lm}(\mathbf{%
\Omega })\,v(r)/r^{\nu }$, with $Y_{lm}(\mathbf{\Omega })$ being the
hyperspherical harmonics.

In the strong coupling regime for $\lambda \geq (l+\nu )^{2}$, (\ref
{eq:MacDonald_asymptotic}) displays, near the origin,~log-periodic
oscillatory behavior.\ we would like to emphasize that while this is usually
considered an \textit{intriguing} property \cite{Camblong:2003mz}, this type
of behavior turns out to be the signature of discrete scale-invariance and,
as we shall see in the next sections, is ever present in many physical
quantities in this and related models. Again, in a comparison with $q$%
-deformed models (with $q$ real), notice that with the natural definition of
a $q$-parameter $q=\mathrm{e}^{\frac{2\pi }{\Theta }},$ the wavefunction
satisfies: 
\begin{equation}
v^{(>)}(qr)=v^{(>)}(r),  \label{qper}
\end{equation}
this $q$-periodic property, is actually rather ubiquitous in the context of $%
q$-deformations and $q$-calculus, as we shall see. In addition, it appears
when studying other features associated to this model, like the presence of
a limit cycle in renormalization group flows. Note that $\left( \ref{qper}%
\right) $ can also be considered as a particular case of self-similarity,
and thus the connection with fractal behavior can be hinted as well.

\section{Spectral behavior: log-periodic oscillations}

In this Section, we shall study the statistical mechanics quantities
associated to spectra such as $\left( \ref{exp}\right) $ and $\left( \ref
{sinh}\right) .$ Namely, partition functions and density of states. We shall
study these spectral functions employing the well-known tool of zeta
functions \cite{Elizalde:zk} (that is to say: an asymptotic study using
Mellin transforms \cite{Flaj}). In this sense, we expect the approach to
have some interest from a methodological point of view. In particular, it
shows how easily the usual framework of the zeta functions employed in
physics -mainly known as zeta regularization or heat kernel techniques \cite
{Elizalde:zk,Vassilevich:2003xt} - has to be enhanced already with
relatively simple quantum mechanical models. Typically, the usual zeta
functions are characterized by a rather constrained meromorphic structure.
In contrast, the models here discussed easily develop an infinite number of
complex poles in the Mellin transform of the spectral function. As we shall
see, these complex poles are intimately related to fractal behavior \cite
{long,Sor}. Recall now that the usual framework employed in heat-kernel
approaches is borrowed from the theory of pseudodifferential operators and
Riemannian geometry. More precisely, in the theory of pseudodifferential
operators ($\Psi $DO) the relation between the heat kernel and zeta
functions is the following. Let $A$ a pseudodifferential operator ($\Psi $%
DO), fulfilling the conditions of existence of a heat kernel and a zeta
function (see, e.g., \cite{Elizalde:zk}). Its corresponding heat kernel is
given by (see \cite{Elizalde:zk}, and references therein): 
\begin{equation}
K_{A}(t)=\mbox{Tr }e^{-tA}={\sum_{\lambda \in \mbox{Spec }A}}^{\prime
}e^{-t\lambda },
\end{equation}
which converges for $t>0$, and where the prime means that the kernel of the
operator has been projected out before computing the trace, and once again
the corresponding zeta function: 
\begin{equation}
\zeta _{A}(s)=\frac{1}{\Gamma (s)}\mbox{Tr }\int_{0}^{\infty
}t^{s-1}\,e^{-tA}\,dt.
\end{equation}
For $t\downarrow 0$, we have the following asymptotic expansion: 
\begin{equation}
K_{A}(t)\sim \alpha _{n}(A)+\sum_{n\neq j\geq 0}\alpha
_{j}(A)\,t^{-s_{j}}+\sum_{k\geq 1}\beta _{k}(A)\,t^{k}\ln t,\quad
t\downarrow 0,
\end{equation}
being: 
\begin{eqnarray}
\alpha _{n}(A) &=&\zeta _{A}(0),\quad \alpha _{j}(A)=\Gamma (s_{j})%
\mbox{Res
}_{s=s_{j}}\zeta _{A}(s),\ \ \mbox{if }s_{j}\notin Z\ \mbox{or }s_{j}>0, \\
\alpha _{j}(A) &=&\frac{(-1)^{k}}{k!}\left[ \mbox{PP }\zeta _{A}(-k)+\left(
1+\frac{1}{2}+\cdots +\frac{1}{k}-\gamma \right) \ \mbox{Res }_{s=-k}\zeta
_{A}(s)\right] ,  \notag \\
\beta _{k}(A) &=&\frac{(-1)^{k+1}}{k!}\,\mbox{Res }_{s=-k}\zeta _{A}(s), 
\notag
\end{eqnarray}

Zeta functions with complex poles have already appeared in the literature.
First, the problem of the study of vibrations in the presence of irregular
boundaries or shapes has already been undertaken as a generalization of
spectral problems with smooth geometries \cite{long}. On the other hand, in
the study of combinatorial structures \cite{Flaj}, Mellin transforms with
complex poles are the most usual and natural object and, in particular,
extremely simple functions in analytic number theory posses this property 
\cite{Flaj}.

Regarding the appearance of poles with imaginary part $s^{*}=\sigma +it,$ we
should only mention that they imply the presence of fluctuations of the type 
$x^{-s^{*}}=x^{-\sigma }\exp \left( it\log x\right) $ in the expansion of
the original function. Very often, regularly spaced poles appear, leading to
a Fourier series in $\log x$. The following (simplified) table gives an idea
of the correspondence between the original function and the singularity
structure of its associated Mellin transform \cite{Flaj}: 
\begin{equation}
\begin{tabular}{|c|c|}
\hline
$f^{*}\left( s\right) $ & $f\left( x\right) $ \\ \hline
Pole at $\xi $ & Term in asymptotic expansion $\approx x^{-\xi }$ \\ \hline
Multiple pole: $\frac{1}{\left( s-\xi \right) ^{k}}$ & Logarithmic factor: $%
\frac{\left( -1\right) ^{k}}{k!}x^{-\xi }\left( \log x\right) ^{k}$ \\ \hline
Complex pole: $\xi =\sigma +it$ & Fluctuations: $x^{-\xi }\exp \left( it\log
x\right) $ \\ \hline
\end{tabular}
\end{equation}

\subsection{Pure geometric growth}

Without losing generality, let us begin by showing the features of the
spectrum $\lambda _{n}=2^{n}$. The associated partition function is: 
\begin{equation}
K(t)=\sum_{n=0}^{\infty }\mathrm{e}^{-2^{n}t}.
\end{equation}
We can construct and associated zeta function as: 
\begin{equation}
\zeta \left( s\right) =\sum_{n=0}^{\infty }2^{-sn}=\frac{1}{1-2^{-s}},
\end{equation}
it is a geometric series and has infinitely many (complex) poles: 
\begin{equation}
s_{k}=\frac{2i\pi k}{\log 2},\quad k=0,\pm 1,\pm 2,...
\end{equation}
with residues $1/\log 2$. Since the corresponding Mellin transform is $%
\Gamma \left( s\right) \zeta \left( s\right) ,$we have a double pole at $0$,
the infinite sequence of complex poles and the poles of the Gamma function
at the negative integers. Then, the asymptotic expansion (see Appendix) is: 
\begin{equation}
K\left( t\right) _{t\rightarrow 0}\sim -\log _{2}\left( t\right) -\frac{%
\gamma }{\log 2}+\frac{1}{2}-Q\left( \log _{2}x\right) +\sum_{n=0}^{\infty }%
\frac{1}{1-2^{n}}\frac{\left( -t\right) ^{n}}{n!}.  \label{asy}
\end{equation}
Where $Q\left( \log _{2}x\right) $ is the contribution from the imaginary
poles: 
\begin{equation}
Q\left( \log _{2}x\right) =\frac{1}{\log 2}\sum_{k\neq 0}\Gamma \left(
s_{k}\right) \exp \left( -2ik\pi \log _{2}x\right) ,
\end{equation}
which is a fluctuating term of order $O(1)$. Then, we have found that the $%
t\rightarrow 0$ expansion for this system contains a term that is a Fourier
series in $\log _{2}x,$ with coefficients of Gamma type. Note that these
oscillations are logarithmic oscillations, typical in discrete-scale
invariant models \cite{Sor}. Notice that $\left| \Gamma \left( \pm 2\pi
i/\log 2\right) \right| =0.545\cdot 10^{-6},$ and, in addition, $\Gamma
\left( s\right) $ has a strong decay while progressing through the imaginary
line. More precisely, recall the complex version of Stirling's formula \cite
{Abra}:
\begin{equation}
\left| \Gamma \left( \sigma +it\right) \right| \sim \sqrt{2\pi }\left|
t\right| ^{\sigma -1/2}\mathrm{e}^{-\pi \left| t\right| /2}.
\end{equation}

Thus, the terms coming from higher poles in the imaginary axes are strongly
damped. Then, the attentive reader may profitably wonder whether the
fluctuations in our models imply that the resulting function is fractal 
\footnote{%
That is, continuous but differentiable nowhere.}. Actually, regarding $%
K\left( t\right) $ it turns out that the great damping in the amplitudes due
to the Gamma function, as explained above, prevents this to be the case \cite
{long}. Therefore, we are dealing with small fluctuations, with tiny \textit{%
wobbles }\cite{Flaj}.

This is not the case of the density of states: 
\begin{equation}
\rho \left( x\right) =\sum_{n=0}^{\infty }\delta \left( x-\lambda
_{n}\right) ,
\end{equation}
since the Mellin transform of the density of states is given by $\zeta
\left( 1-s\right) $, and then the expansion for the density of states is: 
\begin{equation}
\rho \left( x\right) _{x\rightarrow \infty }\sim \frac{x^{-1}}{\log 2}%
\sum_{k=-\infty }^{\infty }\cos \left( 2k\pi \log _{2}x\right) .
\end{equation}
Let us explore some other physically interesting spectral function to learn
more about the role of the poles and its residues. In some contexts, $q$
deformations have proven useful in the study of disordered systems, where is
interesting to understand the conductance, given by: 
\begin{equation}
\left\langle g\left( t\right) \right\rangle =\sum_{n=0}^{\infty }\frac{1}{%
1+\lambda _{n}/t},
\end{equation}
with $\lambda _{n}$ have, in principle, a behavior of the type discussed in
this Section. Once again, we can set $\lambda _{n}=2^{n}$ without losing
generality, and compute the Mellin transform: 
\begin{equation}
g\left( s\right) =\int_{0}^{\infty }\left\langle g\left( t\right)
\right\rangle t^{s-1}dt=\frac{\pi }{\sin \pi \left( s-1\right) }\frac{1}{%
1-2^{s}},
\end{equation}
and taking into account all the poles in the Mellin transform, we obtain: 
\begin{equation}
\left\langle g\left( t\right) \right\rangle _{t\rightarrow \infty }\sim \log
_{2}t+\frac{1}{2}+P\left( \log _{2}t\right) +\sum_{k=1}^{\infty }\frac{%
\left( -1\right) ^{k}}{1-2^{k}}t^{-k},
\end{equation}
where $P\left( \log _{2}t\right) $ is the periodic function coming from the
complex poles $s_{k}=\frac{2\pi ik}{\log 2}$: 
\begin{eqnarray}
P\left( \log _{2}t\right) &=&\frac{1}{\log 2}\sum_{k\in Z}\frac{\pi }{\sin
\pi \left( s_{k}-1\right) }\exp \left( -2ik\pi \log _{2}x\right) \\
&=&-\frac{2\pi }{\log 2}\sum_{k=1}^{\infty }\frac{\sin 2\pi k\log _{2}t}{%
\sinh \left( \frac{2\pi ^{2}k\log _{2}t}{\log 2}\right) }.  \notag
\end{eqnarray}

Once again, the important point is that the function $\frac{\pi }{\sin \pi
\left( s-1\right) }$ exhibits an exponential decrease along vertical lines.
Then, the contribution of the periodic function is truly small. This can be
readily appreciated from the $\sinh $ factor, that severely damps the
contribution of higher poles. More precisely, the function satisfies $\left|
P\left( t\right) \right| \leq 7.8\cdot 10^{-12}.$ Nevertheless, it is plain
that they can have a more considerable effect (for example, by decreasing
the value of $q$, that we have set to $\frac{1}{2},$ just to give precise
numerical values). In any case, it is interesting to point out that even
such minute fluctuations have lead to discrepancies in analytic studies of
algorithms \cite{Flaj}.

\subsection{$q$-deformed growth}

In the previous Section we have studied in detail the case of a pure
geometric growth. Let us now pay attention to the very well-known case of a $%
q$-deformed spectrum. From a practical point of view, numbers are
substituted by $q$-numbers: 
\begin{equation}
n\rightarrow \left[ n\right] _{q}=\frac{1-q^{n}}{1-q}.
\end{equation}
Note that the resulting spectrum exhibits exponential growth for the
high-level eigenvalues and linear growth for the low-lying eigenvalues: 
\begin{eqnarray}
\lim_{n\rightarrow \infty }\left[ n\right] _{q}\left( q-1\right) &\sim &%
\mathrm{e}^{n\log q} \\
\lim_{n\rightarrow 0}\left[ n\right] _{q}\left( q-1\right) &\sim &n\log q. 
\notag
\end{eqnarray}
The effect of this spectrum -instead of a purely exponential one- in the
meromorphic structure of the associated zeta function is not evident a
priori.

Interestingly enough, it turns out that there already exist some few
interesting works dealing with $q$ deformed zeta functions \cite
{qzeta2,Chered,qzeta3}. Consider, for example, \cite{Chered}: 
\begin{equation}
\zeta _{q}\left( s\right) =\sum_{n=1}^{\infty }\frac{q^{sn}}{\left[ n\right]
_{q}^{s}}.  \label{zeta1}
\end{equation}
The meromorphic continuation to all $s$ can be easily obtained by
application of the binomial expansion $\left( 1-q^{n}\right)
^{-s}=\sum_{r=0}^{\infty }\binom{s+r-1}{r}q^{nr}$ \cite{qzeta2}: 
\begin{equation}
\zeta _{q}\left( s\right) =\left( 1-q\right) ^{s}\sum_{r=0}^{\infty }\binom{%
s+r-1}{r}\frac{q^{s-1+r}}{1-q^{s-1+r}}.
\end{equation}
Thus, the set of poles is $s_{k,n}=-k+\frac{2\pi in}{\log q}$ with $%
k=0,-1,-2,...$ and $n$ an integer. Hence, as expected, we are lead to
complex poles and thus to logarithmic oscillations and, eventually, to
fractal behavior as discussed previously. Nevertheless, in comparison with
the previous Section, the singularity pattern is manifestly much richer.
Since there are many possible $q$ deformations, it is worth to introduce,
following \cite{qzeta3}, a generic two-variable zeta function: 
\begin{equation}
f_{q}\left( s,t\right) =\sum_{n=1}^{\infty }\frac{q^{nt}}{\left[ n\right]
_{q}^{s}}\quad \text{with }\left( s,t\right) \in {\mathbb{C}}^{2},
\end{equation}
and proceed identically. Namely, binomial expansion and change of the order
of summation \cite{qzeta3}: 
\begin{equation}
f_{q}\left( s,t\right) =\left( 1-q\right) ^{s}\sum_{r=0}^{\infty }\binom{%
s+r-1}{r}\frac{q^{t+r}}{1-q^{t+r}}.  \label{general}
\end{equation}
In \cite{qzeta3}, it is argued that $t=s-1$ provides the ''right'' $q$%
-deformation of Riemann's zeta function, instead of the case $t=s$ presented
above and discussed in \cite{qzeta2}. Certainly, we are led to: 
\begin{equation}
f_{q}\left( s,s-1\right) =\left( 1-q\right) ^{s}\left( \frac{q^{s-1}}{%
1-q^{s-1}}+s\frac{q^{s}}{1-q^{s}}+\frac{s\left( s+1\right) }{2}\frac{q^{s+1}%
}{1-q^{s+1}}+...\right) ,
\end{equation}
and consequently the poles are simple at $1+\frac{2\pi in}{\log q}$ with $%
n\in {\mathbb{Z}}$ and $j+\frac{2\pi ik}{\log q}$ with $j,k\in $ ${\mathbb{Z}%
}$ and $j\leq 0,k\neq 0.$

Therefore, in comparison with $\left( \ref{zeta1}\right) ,$ we avoid the
poles at negative integer values and at the origin. In addition, the
corresponding values at these points approach the known ones, when $%
q\rightarrow 1$. Furthermore, the convergence to $\zeta \left( s\right) $
when $q\rightarrow 1$ for any $s$ is obtained. This is also possible for $%
\left( \ref{zeta1}\right) $, but up to some terms in $q$ \cite{qzeta2}. With
the previous information, it is straightforward now to study the asymptotic
behavior of the associated partition function: 
\begin{equation}
K\left( t\right) =\sum_{n=1}^{\infty }q^{-n}\mathrm{e}^{-\frac{q^{n}}{\left[
n\right] _{q}}t},
\end{equation}
\begin{eqnarray}
K\left( t\right) _{t\rightarrow 0} &\sim &\frac{\left( 1-q\right) }{t}%
\sum_{k=-\infty }^{\infty }\Gamma \left( 1+\frac{2\pi ik}{\log q}\right)
\cos \left( 2\pi k\frac{\log x}{\log q}\right)  \\
&&+\sum_{n=0}^{\infty }\zeta _{q}\left( -n\right) \frac{\left( -t\right) ^{n}%
}{n!}  \notag \\
&&+\sum_{j=0}^{\infty }\sum_{k\neq 0}t^{j}\Gamma \left( s_{j,k}\right)
\left( s_{j,k}\right) _{j}\left( 1-q\right) ^{s_{j,k}}\cos \left( 2\pi k%
\frac{\log x}{\log q}\right) .  \notag
\end{eqnarray}
As mentioned, the zeta values $\zeta _{q}\left( -n\right) $ are finite and
its value well-known \cite{qzeta3}. The $s_{j,k}$ denote the poles and $%
\left( s_{j,k}\right) _{j}$ the Pochhammer symbol (also known as rising
factorial) \cite{Abra}.

However, as explained in the Introduction, we are interested in spectrum of
the type $\lambda _{n}=\sinh \left( n\log q\right) $. This directs our
interest to $\left( \ref{general}\right) $ with $t=s/2$. Thus:
\begin{eqnarray}
\zeta _{q}\left( s\right)  &=&\sum_{n=1}^{\infty }\frac{q^{ns/2}}{\left[
n\right] _{q}^{s}}=\left( \frac{q-1}{2}\right) ^{s}\sum_{n=1}^{\infty }\sinh
^{-s}\left( \frac{n\log q}{2}\right)  \\
&=&\left( 1-q\right) ^{s}\sum_{r=0}^{\infty }\binom{s+r-1}{r}\frac{q^{s/2+r}%
}{1-q^{s/2+r}}.  \notag
\end{eqnarray}
This zeta function also posses many interesting theoretical properties. In
particular, the mathematical relevance -in the theory of special functions
and in harmonic analysis- of the transformation $x^{s}\rightarrow \sinh ^{s}x
$ is discussed in \cite{Chered}. Additionally, it actually appears in a
natural way in physical applications. In random matrix models for example, a
model with the usual Gaussian potential but with $\left( x_{i}-x_{j}\right)
^{2}\rightarrow \sinh ^{2}\left( (x_{i}-x_{j})/2\right) $ as a correlation
factor, appears in Chern-Simons theory \cite{Marino:2002fk,Tierz:2002jj}.

The set of poles is $2a+\frac{2\pi ib}{\log q}$ with $a,b\in {\mathbb{Z}}$
and $a\leq 0.$ As usual, the set of poles allows to obtain the rich
asymptotic behavior of the density of states or the partition function. In
any case, note that the contribution of the additional poles is subleading
against the rightmost line of complex poles (and in comparison with the
purely geometric growth case).

\section{QFT on a discretely self-similar spacetime: one-loop effects.}

Now, we shall show how $q$-deformed zeta functions (and the corresponding
heat kernels) and the above mentioned $q$-periodic functions, naturally
appear in essentially the same way as ordinary zeta functions do in zeta
regularization/heat-kernel studies \cite{Elizalde:zk,Vassilevich:2003xt}.
Namely, in the study of one-loop effects on a curved space-time background.

For this, we focus our attention on a certain type of space-times that
appear in the study of critical phenomena in gravitational collapse .
Indeed, from Choptuik's groundbreaking work \cite{Choptuik:1992jv}, much
attention has been recently devoted to metrics with a discrete
self-similarity \cite{Gundlach:2002sx,Gundlach:2003pg}. A spacetime is
discretely selfsimilar if there exists a discrete diffeomorphism $\Phi $ and
a real constant $\Delta $ such that: 
\begin{equation}
\Phi ^{*}g_{ab}=\mathrm{e}^{-2\Delta }g_{ab},
\end{equation}
where $\Phi ^{*}g_{ab}$ is the pull-back of $g_{ab}$ under the
diffeomorphism $\Phi .$

Notice that, in contrast with the continuous case, this definition does not
introduce an homothetic vector field $\xi .$ The parameter is essentially
the analogous of $\log q$ in our discussion. In Schwarzchild-like
coordinates, the spacetime line element reads:
\begin{equation}
ds^{2}=-\alpha ^{2}\left( r,t\right) dt^{2}+a^{2}\left( r,t\right)
dr^{2}+r^{2}d\Omega ^{2},  \label{DSS}
\end{equation}
where the coefficients satisfy the property: 
\begin{equation}
\alpha \left( \mathrm{e}^{\Delta }r,\mathrm{e}^{\Delta }t\right) =\alpha
\left( r,t\right) \text{ and }a\left( \mathrm{e}^{\Delta }r,\mathrm{e}%
^{\Delta }t\right) =a\left( r,t\right) ,  \label{DSS2}
\end{equation}
the by-now familiar $q$-periodic property. In spite of the interest they
have generated, these metrics are still much less studied than the usual
continuously self-similar spacetimes, for example. Let us also briefly
discuss the $q$-periodic property (\ref{DSS2}).

\subsection{On $q$-periodic functions}

Recall that when we deal with a noncommutativity of the type $xy=qyx$
(Manin's \textit{quantum plane), }then ordinary derivatives are substituted
by $q$-derivatives \cite{Majid:kd,Eck}: 
\begin{equation}
\partial _{x}^{\left( q\right) }f\left( x;y;...\right) \equiv \frac{f\left(
qx;y;...\right) -f\left( x;y;...\right) }{\left( q-1\right) x}.  \label{qder}
\end{equation}
Notice how the $q$ derivative measures the rate of change with respect to a
dilatation of the argument, instead of the translation of the usual
derivative. Then, from the previous expression, it is manifest that a $q$%
-periodic function satisfies: 
\begin{equation}
\partial _{L}^{\left( q\right) }g\left( L\right) =0  \label{qperconst}
\end{equation}
Actually, the unique solution to this equation is a $q$-periodic function
and, of course, as in the classical case, a constant. Therefore, it can be
said that a $q$-periodic function plays, at the level of $q$-calculus, the
role of a constant in ordinary (commutative) calculus.

The connection with complex dimensions \cite{long,Sor,Flaj} can be easily
obtained. We consider the Mellin transform of the $q$ periodic function: 
\begin{equation}
h(s)\equiv \int_{0}^{\infty }g\left( x\right) x^{s}dx.
\end{equation}
Taking into account the following property of Mellin transforms (see \cite
{Flaj} for example): 
\begin{equation}
\int_{0}^{\infty }g\left( qx\right) x^{s}dx=q^{-s}h(s),
\end{equation}
and considering the $q$-periodic property $\left( \ref{qperiodic}\right) $: 
\begin{equation}
q^{-s}h(s)=h\left( s\right) \Rightarrow s_{k}=\frac{2\pi in}{\log q},\quad
n\in {\mathbb{Z}.}
\end{equation}
Thus, the Mellin transform of a $q$ periodic function contains infinitely
many complex poles. As we already know, this implies that the $q$ periodic
function may be fractal. As we have already seen in the other sections, this
depends on the precise form of the function itself. Namely, on the residue
corresponding to the poles.

Equivalently, one may argue that any log-periodic term satisfy the
restriction imposed by $\left( \ref{qperiodic}\right) $, so one can
construct any suitable combination, such as: 
\begin{equation}
f\left( x\right) =\sum_{n=1}^{k}a_{n}\sin \left( 2\pi b_{n}\frac{\log x}{%
\log q}\right) ,
\end{equation}
with rather generic coefficients $a_{n}$ and $b_{n}$ Nevertheless, we are
still limiting the finest scale possible, since the sum stops at $n=k$ and
thus the function is not a genuine fractal, albeit the oscillatory pattern
is certainly much richer than a single log-periodic term. But we can
consider a full Fourier series and take $k\rightarrow \infty $. Consider for
example, $a_{n}=n^{-\gamma }$ with $\gamma >1;$ then: 
\begin{equation}
f\left( x\right) =\sum_{n=1}^{\infty }\frac{1}{n^{\gamma }}\sin \left( 2\pi n%
\frac{\log x}{\log q}\right) ,  \label{general2}
\end{equation}
It is well-known that low enough values of $\gamma $ lead to fractal
behavior. Therefore, both relatively simple log-periodic patterns that lead
to smooth functions and much more fluctuating patterns leading to
fractal functions (see Figures below) are $q$-periodic.

\begin{figure}[tph]
\centering
\includegraphics{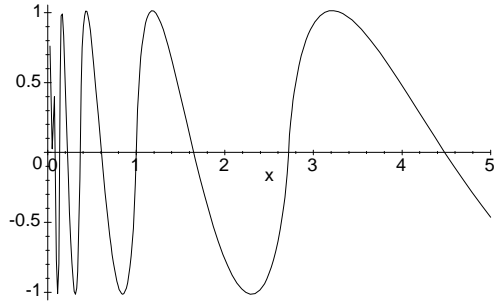}
\caption{$a_{n}=\frac{1}{n^{2}},$ $k=200$ and $b_{n}=n$}
\end{figure}

\begin{figure}[tph]
\centering
\includegraphics{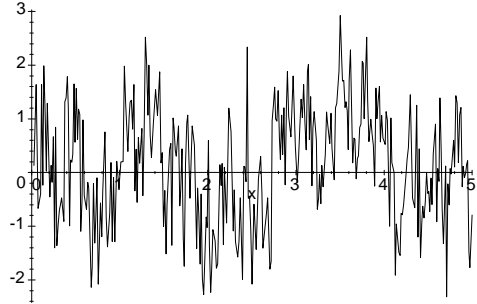}
\caption{$a_{n}=\frac{1}{n^{0.3}},$ $k=200$ and $b_{n}=n^{2}$}
\end{figure}

Another interesting aspect in the study of models characterized by a
discrete scale invariance, is the appearance of a limit cycle in a
renormalization group analysis. A limit cycle is a 1-parameter family of
coupling constants that is closed under the RG flow. The necessary
adimensional parameter corresponds to the angle $0<\theta <2\pi $ and the
scale invariance is discrete. In the several models where a cyclic RG behavior 
has been found, it turns out that the couplings return to their initial values 
after a finite RG time $\lambda :$%
\begin{equation}
g\left( \mathrm{e}^{\lambda }L\right) =g\left( L\right)  \label{qperiodic}
\end{equation}
where $L$ is the RG length scale. Note that many other interesting physical
quantities are also characterized by this $q$-periodic property \cite
{LeClair:2002ux}.

Needless to say, from a physical standpoint one would like, perhaps, to rule
out the fractal case, since this implies the non-differentiability of $%
g\left( L\right) $. Further work to clarify this would be desirable,
especially taking into account that one can construct intermediate cases,
where the resulting function is neither smooth nor fractal, but begins to
develop a countable set of singularities \cite{GluzSor}.

\subsection{Discretely self-similar space-time as a quantum group}

As we have just seen, discrete selfsimilarity leads to coefficients of the
metric which are the $q$-periodic functions previously discussed. So,
already at the classical level, the presence of the discrete self-similarity
may imply a very low differentiability of the space-time. This may not
necessarily jeopardize the validity of such an spacetime \cite{low}.

This happens already at the level of the classical geometry, but what about
one-loop effects ? Let us look at the heat kernel. To begin with, one can
not directly apply the typical methods \cite{Elizalde:zk,Vassilevich:2003xt}%
. Consider, the symbol of a Laplacian operator on a Riemannian manifold: 
\begin{equation}
a\left( x,\xi \right) ={-g^{ij}(x)\xi _{i}\xi _{j}+B^{i}(x)\xi _{i}+C(x).}
\end{equation}

Then, as in \cite{Vassilevich:2003yz}, a signal that the usual framework can
not be applied is that the metric (\ref{DSS})-(\ref{DSS2}) leads to an
oscillating symbol. The difference with \cite{Vassilevich:2003yz} is that in
the present case the oscillatory behavior is log-periodic.

Now, it is manifest that the existence of the $\Delta $ parameter, the
equivalent of $\log q$, will lead to functional behavior in the spectra of
the Laplacian. For definiteness, let us briefly consider a very simple
particular example. Consider the one-dimensional case: 
\begin{equation}
ds^{2}=g(x)dx^{2},
\end{equation}
where $g(x)$ is of course $q$-periodic. Then, solving for the spectra of the
Laplacian implies: 
\begin{equation}
g(x)\phi ^{\prime \prime }\left( x\right) =E_{n}\phi \left( x\right) ,
\end{equation}
And the $q$-periodic property of $g(x)$ readily implies that $\phi \left(
qx\right) =q^{-2}\phi \left( x\right) .$ Thus, the self-similarity of the
wavefunction is an automatic consequence. Note that this already implies
self-similarity of the full heat-kernel $K(x,y,t,D)=\left\langle x\left|
\exp \left( -tD\right) \right| y\right\rangle $. Of course, such a
wavefunction can now be expanded in a log-periodic Fourier series. However,
it is simpler to notice that a function satisfying $\phi \left( qx\right)
=q^{-2}\phi \left( x\right) $ can also be given by: 
\begin{equation}
\phi \left( x\right) =\sum_{n=-\infty }^{\infty }\left( 1-\exp \left(
iq^{n}x\right) \right) q^{2n}  \label{qosc}
\end{equation}
Although we have written the complex wavefunction, it is clear that these
are usual trigonometric oscillations with the only novelty that the
frequencies of the oscillations are of the type $\omega _{n}=q^{n}$ . That
is to say, with the geometric growth we have been discussing in this paper.
This obviously translates into a geometric growth for the energy eigenvalues
and thus, to a heat kernel and zeta functions of the type discussed in the
previous Section.

Notice that all the previous discussions on the appearance of fractal
behavior are now relevant as we already know that this novel behavior in the
short-time asymptotics of the trace of the heat-kernel, is of very small
amplitude. Recall that this expansion gives the large mass expansion of the
effective action \cite{Vassilevich:2003xt}.

Therefore, we have seen that, at a quantum level, discretely self-similar
spacetimes are very much related to quantum groups \footnote{%
Recall that we can consider a quantum group as a manifold, and its spectral
analysis is of the type here discussed \cite{qzeta2}.}. It seems appealing
that the study of critical phenomena in gravitational collapse, that
naturally leads to the discrete symmetry characterized by $\Delta ,$ might
be related with quantum group symmetries. This is so because a quantum group
symmetry is expected to be a key ingredient in quantum gravity \cite
{Amelino-Camelia:2003xp} and, consequently, in black hole physics. Thus, it
seems appealing that Choptuik's result could be interpreted as a signature
of a quantum group symmetry and thus, a full-fledged quantum gravity effect.

\section{Conclusions and Outlook}

We have focussed on analytical aspects of $q$-deformations with $q$ real.
This has been done by simple comparison with discrete scale invariance \cite
{Sor}, a well-known topic in many statistical physics applications (although
often appearing in a scattered fashion in the literature, in spite of
reviews such as \cite{Sor}). We have seen that some models of conformal
quantum mechanics, in the strong coupling regime an after regularization,
also lead to the same behavior. In this sense, once this connection is
realized, the properties in this regime are not intriguing but rather
expected ones.

Incidentally, the tight connections between discrete scale invariance and
fractal geometry puts into evidence the relationship between such models, $q$%
-deformations and fractal behavior. In addition, the functional behavior in
the spectra, which is a consequence of the $q$-deformation or, more
generically, of the presence of a discrete scale symmetry, naturally leads
to associated spectral functions (such as zeta functions and trace of heat
kernels) with interesting behavior. While such functions have appeared in
mathematical literature, they have not been considered in zeta
regularization \cite{Elizalde:zk,Vassilevich:2003xt} and in this context,
they show many interesting novel features. Mainly, a very rich meromorphic
structure, which is of course consequence of the discrete scale invariance.
This seems to be interesting, but the main reason is not one of just
mathematical generality, as we have also shown that quantum field theory on
discretely self-similar space-times exactly requires such an extended
framework.

Precisely, regarding the discretely self-similar space-times, there seems to
be several interesting open questions. On a mathematical level, it seems
interesting to further study the low differentiability properties of some of
the space-times included in the generic definition (\ref{DSS})-(\ref{DSS2}).
This may be done along the lines of \cite{low} for example. From a more
physical point of view, we hope to have shown that quantum field theory on
these backgrounds is an interesting and rather unexplored problem. To begin
with, there is the connection with quantum group symmetries and the possible
meaning in quantum gravity, perhaps even also the comparison with
developments on noncommutative field theory. If one is more interested in
technical issues, we have already seen that in a heat-kernel/zeta functions
approach, novel behavior appears, so a more careful study of this aspect may
be worth as well, maybe in comparison with heat-kernels on noncommutative
spaces \cite{Vassilevich:2003yz}.

\medskip

\bigskip

\textbf{Acknowledgments}

\medskip

The author is grateful to Jean-Pierre Eckmann, Sebastian de Haro, Emilio
Elizalde and Michel Lapidus for comments and discussions.


\appendix

\newpage

\section{Mellin transforms and zeta functions: complex poles}

To complement the information provided in the Introduction, we just quote
here the theorem that is implicitly used in the computations of the
asymptotic behavior of the spectral functions. We note in passing that not
only a geometric growth of the spectral sequence $\left\{ \lambda
_{n}\right\} $ leads to complex poles. An arithmetic sequence with varying
amplitudes (degeneracies) may exhibit the same behavior. That is, Dirichlet
series such as: 
\begin{equation}
L(s)=\sum_{n}\frac{\upsilon _{2}\left( n\right) }{n^{-s}},
\end{equation}
where $\upsilon _{2}\left( n\right) $ denotes the exponent of $2$ in the
prime number decomposition of the integer $n,$ has complex poles \cite{Flaj}%
. The theorem is as follows.

\begin{theorem}
Let $f(x)$ be continuous in $\left] 0,\infty \right[ $ with Mellin transform 
$f^{*}\left( s\right) $ having a non-empty fundamental strip $\left\langle
\alpha ,\beta \right\rangle $.

$\left( i\right) $ Assume that $f^{*}\left( s\right) $ admits a meromorphic
continuation to the strip $\left\langle \gamma ,\beta \right\rangle $ with
some $\gamma <\alpha $ with a finite number of poles there, and is analytic
on $\Re \left( s\right) =\gamma .$

Assume also that there exists a real number $\eta \in \left( \alpha ,\beta
\right) $ such that

\begin{equation}
f^{*}\left( s\right) =O\left( \left| s\right| ^{-r}\right) \text{ with }r>1,
\end{equation}

when $\left| s\right| \rightarrow \infty $ in $\gamma \leq $ $\Re \left(
s\right) \leq \eta .$ If $f^{*}\left( s\right) $ admits the singular
expansion for $s\in $ $\left\langle \gamma ,\alpha \right\rangle $ 
\begin{equation}
f^{*}\left( s\right) \asymp \sum_{\left( \xi ,k\right) \in A}d_{\xi ,k}\frac{%
1}{\left( s-\xi \right) ^{k}},
\end{equation}

then an asymptotic expansion of $f(x)$ at $0$ is 
\begin{equation}
f(x)=\sum_{\left( \xi ,k\right) \in A}d_{\xi ,k}\left( \frac{\left(
-1\right) ^{k-1}}{\left( k-1\right) !}x^{-\xi }\left( \log x\right)
^{k}\right) +O\left( x^{-\gamma }\right) .
\end{equation}
\end{theorem}

There is also an analogous statement for an asymptotic expansion of $f(x)$
at $\infty .$ Proofs can be found in \cite{Flaj}. To conclude, we just want
to point out that in many cases, $f^{*}\left( s\right) $ is meromorphic in a
complete left or right half plane, and then a complete asymptotic expansion
for $f^{*}\left( s\right) $ results. Such an expansion can be convergent or
divergent. If divergent, the expansion is then only asymptotic. If
convergent it may represent the function exactly. In general, we need a fast
enough and uniform decrease of $f^{*}\left( s\right) $ along vertical lines.
The example treated in Section $2$ for example, satisfies this property and
then, the representation is exact. More details can be found in \cite{Flaj}.


\begin{thebibliography}{99}
\bibitem{Camblong:2000ec}  H.~E.~Camblong, L.~N.~Epele, H.~Fanchiotti and
C.~A.~Garcia Canal, 
Phys.\ Rev.\ Lett.\ \textbf{85}, 1590 (2000) [arXiv:hep-th/0003014]; 
Annals Phys.\ \textbf{287}, 14 (2001) [arXiv:hep-th/0003255]; 
Annals Phys.\ \textbf{287}, 57 (2001) [arXiv:hep-th/0003267].


\bibitem{Camblong:2003mz}  H.~E.~Camblong and C.~R.~Ordonez, 
Phys.\ Rev.\ D \textbf{68}, 125013 (2003) [arXiv:hep-th/0303166]. 


\bibitem{Skorik:1993ce}  S.~Skorik and V.~Spiridonov, 
Lett.\ Math.\ Phys.\ \textbf{28}, 59 (1993) [arXiv:hep-th/9304107]; 
V.~Spiridonov, 
Phys.\ Rev.\ A \textbf{52}, 1909 (1995) [arXiv:quant-ph/9601030]; 
I.~Loutsenko and V.~Spiridonov, 
[arXiv:solv-int/9909022]. 

\bibitem{Braaten:2003eu}
E.~Braaten and H.~W.~Hammer,
Phys.\ Rev.\ Lett.\  {\bf 91}, 102002 (2003)
[arXiv:nucl-th/0303038].


\bibitem{Glazek:1993rc}  S.~D.~Glazek and K.~G.~Wilson, 
Phys.\ Rev.\ D \textbf{48}, 5863 (1993). 
Phys.\ Rev.\ Lett.\ \textbf{89}, 230401 (2002) [arXiv:hep-th/0203088]; %
[arXiv:cond-mat/0303297].


\bibitem{LeClair:2002ux}  A.~LeClair, J.~M.~Roman and G.~Sierra, 
Phys.\ Rev.\ B \textbf{69}, 20505 (2004) [arXiv:cond-mat/0211338]; %
Nucl. Phys. \textbf{B} 675, 584 (2003) 
[arXiv:hep-th/0301042]; 
Nucl.\ Phys.\ B \textbf{700}, 407 (2004) [arXiv:hep-th/0312141]. 


\bibitem{Macfarlane:dt}  A.~J.~Macfarlane, 
J.\ Phys.\ A \textbf{22}, 4581 (1989); 
L.~C.~Biedenharn, 
J.\ Phys.\ A \textbf{22}, L873 (1989). 


\bibitem{Majid:kd}  S.~Majid, \emph{Foundations Of Quantum Group Theory,}
Cambridge, UK: Univ. Pr. (1995) 607 p%


\bibitem{deAlfaro:1976je}  V.~de Alfaro, S.~Fubini and G.~Furlan, 
Nuovo Cim.\ A \textbf{34}, 569 (1976). 

\bibitem{Case}  K. M. Case, Phys. Rev. \textbf{80}, 797 (1950).

\bibitem{Sor}  D. Sornette, Phys. Rept. \textbf{297,} 239 (1998). Expanded
version in [arXiv:cond-mat/9707012]


\bibitem{Fichtmuller:1995dt}  M.~Fichtmuller, A.~Lorek and J.~Wess, 
Z.\ Phys.\ C \textbf{71}, 533 (1996) [arXiv:hep-th/9511106]. 

\bibitem{Efimov}  V. Efimov, Sov. J. Nucl. Phys. \textbf{12}, 589 (1971);
Comments Nucl. Part. Phys. \textbf{19}, 271 (1990).


\bibitem{Wilson:1970ag}  K.~G.~Wilson, 
Phys.\ Rev.\ D \textbf{3}, 1818 (1971). 


\bibitem{Bedaque:1998kg}  P.~F.~Bedaque, H.~W.~Hammer and U.~van Kolck, 
Phys.\ Rev.\ Lett.\ \textbf{82}, 463 (1999) [arXiv:nucl-th/9809025]. 


\bibitem{Braaten:2002jv}  E.~Braaten and H.~W.~Hammer, 
Phys.\ Rev.\ A \textbf{67}, 042706 (2003) [arXiv:cond-mat/0203421]; 
Phys.\ Rev.\ Lett.\ \textbf{87}, 160407 (2001) [arXiv:cond-mat/0103331]. 


\bibitem{Morozov:2003ik}  A.~Morozov and A.~J.~Niemi, 
Nucl.\ Phys.\ B \textbf{666}, 311 (2003) [arXiv:hep-th/0304178]. 

\bibitem{CoonHol}  S. A. Coon and B. R. Holstein, Am. J. Phys. \textbf{70},
513 (2002) [arXiv:quant-ph/0202091]


\bibitem{Elizalde:zk}  E.~Elizalde, \emph{Ten Physical Applications Of
Spectral Zeta Functions}, Lect.\ Notes Phys.\ \textbf{M35}, 1 (1995); 
K.~Kirsten, \emph{Spectral Functions In Mathematics And Physics},
Chapman\&Hall/CRC Press, Boca Raton, FL, 2001.%

\bibitem{Flaj}  R. Sedgewick and P. Flajolet, \emph{An introduction to the
Analysis of Algorithms}, Addison-Wesley, 1996

\bibitem{Vassilevich:2003xt}  D.~V.~Vassilevich, Phys.Rept. \textbf{388},
279 (2003) 
[arXiv:hep-th/0306138]; 
A.~A.~Bytsenko, G.~Cognola, L.~Vanzo and S.~Zerbini, 
Phys.\ Rept.\ \textbf{266}, 1 (1996) [arXiv:hep-th/9505061]. 

\bibitem{long}  M. L. Lapidus and M. van Frankenhuysen, \emph{Fractal
Geometry and Number Theory (Complex dimensions of fractal strings and zeros
of zeta functions)}, Research Monograph, Birkh\"{a}user, Boston, 2000.

\bibitem{Abra}  M. Abramowitz and I.A. Stegun (Eds.). \emph{Handbook of
Mathematical Functions with Formulas, Graphs, and Mathematical Tables, 9th
printing.} New York: Dover, 1972.

\bibitem{qzeta2}  K. Ueno and M. Nishizawa, \emph{Quantum groups and
zeta-functions}, Proceedings of the 30-th Karpatz Winter School ''Quantum
Groups: Formalism and Applications'' (1995), 115-126 (Polish Scientific
Publishers PWN) [arXiv:hep-th/9408143] 

\bibitem{qzeta3}  M. Kaneko, N. Kurokawa and M. Wakayama, Kyushu J. Math. 57
(2003) 175-192 [arXiv:math.QA/0206171]

\bibitem{Chered}  I. Cherednik, Sel. math., New Ser. 7, 1-44 (2001)
[arXiv:math.QA/9804099]

\bibitem{Marino:2002fk}
M.~Marino,
Commun.\ Math.\ Phys.\  {\bf 253}, 25 (2004)
[arXiv:hep-th/0207096].

\bibitem{Tierz:2002jj}  M.~Tierz, 
Mod.\ Phys.\ Lett.\ A \textbf{19}, 1365 (2004) [arXiv:hep-th/0212128]. 


\bibitem{Choptuik:1992jv}  M.~W.~Choptuik, 
Phys.\ Rev.\ Lett.\ \textbf{70}, 9 (1993). 


\bibitem{Gundlach:2002sx}  C.~Gundlach, 
Phys.\ Rept.\ \textbf{376}, 339 (2003) [arXiv:gr-qc/0210101]. 


\bibitem{Gundlach:2003pg}  C.~Gundlach and J.~M.~Martin-Garcia, 
Phys.\ Rev.\ D \textbf{68}, 064019 (2003) [arXiv:gr-qc/0306001] 

\bibitem{Eck}  J.P. Eckmann and A. Erzan, Phys. Rev. Lett. \textbf{78}, 3241
(1997).

\bibitem{GluzSor}  S. Gluzman and D. Sornette, Phys. Rev. E 65, 036142 (2002)

\bibitem{low}  C.J.S. Clarke, \emph{The Analysis of Space-Time
Singularities, }Cambridge University Press, 1993.


\bibitem{Vassilevich:2003yz}  D.~V.~Vassilevich, 
Lett.\ Math.\ Phys.\ \textbf{67}, 185 (2004) [arXiv:hep-th/0310144]. 


\bibitem{Amelino-Camelia:2003xp}  G.~Amelino-Camelia, L.~Smolin and
A.~Starodubtsev, 
Class.\ Quant.\ Grav.\ \textbf{21}, 3095 (2004) [arXiv:hep-th/0306134]. 

\end{thebibliography}
\end{document}